# An efficient and effective Decentralized Anonymous Voting System

Wei-Jr Lai
National Taiwan University
r05922108@cmlab.csie.ntu.edu.tw

Ja-Ling Wu
National Taiwan University
wjl@cmlab.csie.ntu.edu.tw

*Abstract*—A trusted electronic election system requires that all the involved information must go public, that is, it focuses not only on transparency but also privacy issues. In other words, each ballot should be counted anonymously, correctly, and efficiently. In this work, a lightweight E-voting system is proposed for voters to minimize their trust in the authority or government. We ensure the transparency of election by putting all message on the Ethereum blockchain, in the meantime, the privacy of individual voter is protected via an efficient and effective ring signature mechanism. Besides, the attractive self-tallying feature is also built in our system, which guarantees that everyone who can access the blockchain network is able to tally the result on his own, no third party is required after voting phase. More importantly, we ensure the correctness of voting results and keep the Ethereum gas cost of individual participant as low as possible, at the same time. Clearly, the pre-described characteristics make our system more suitable for large-scale election.

## I. INTRODUCTION

Although some countries have begun to use electronic voting for national scale election[estonia-voting], there is still no suitable, trusted, efficient electronic voting system for people because it requires many contradictory properties. The election needs one or more authority for both authentication and protect the privacy of participant, however, it is difficult for voter to believe in the government or authority that will always follow the rules or never get hacked. There have been many studies and discussion on elections[voting-emp0], [voting-emp1] open all ballot to ensure transparent, then permute for anonymity, but it is still centralized and tally phase is time-consuming. [voting-emp2] proposed Open Vote network protocol, it is decentralized, anonymous and transparent. But it cannot tally the result even only one voter doesn't cast his ballot, which make their scheme only suitable for small-scale voting. With the development of blockchain, some work has discuss, electronic voting can use it as immutable, public bulletin board. [voting-emp3] run Open Vote protocol on Ethereum smart contract, make whole process more convenient. [voting-emp4,5] build their voting system based on blockchain. [voting-emp4] require all voter open their mask after voting, which property cause the same problem as [voting-emp2,3], and it is obviously unreasonable to force each voter to pay a deposit before voting. [voting-emp5] proposed a simple and efficient scheme, and introduced multiple authority to protect privacy of voter. But it is centralized and has some information that is not completely transparent.

## II. BACKGROUND

### A. Ring signature

The most important element in the electronic election is Anonymous. The relationship between voter and their ballot must not be revealed. Thus, we need a special digital signature scheme called ring signature[ring1], which was first proposed by Ron Rivest, Adi Shamir, and Yael Tauman in 2001. Ring signature has the property that a signer in a particular group can signs the message as group member, and verifier cannot distinguish the identity of signer. We can simply take this method to make voter sign the ballot anonymously but there will be another problem: a voter can cast their ballot twice. This is equivalent to double-spending issue on blockchain network. Thus, we adopt one-time ring signatures[ring2] proposed by Nicolas van Saberhagen, which ensures that a voter with one key-pair can only signs once, but can still sign as a particular group. We describe the one-time ring signature as following:

Common parameters are:
q: a prime number
E: an elliptic curve equation
G: a base point
l: a prime order of the base point
$H_s$: a cryptographic hash function $\{0,1\}^* \to F_q$
$H_p$: a cryptographic hash function $E(F_q) \to E(F_q)$

- Generation

Assume there are a public key set $\{ P_i \mid i \in [1, n] \}$, and a signer owns a private key $x_s$ corresponding to a public key $P_s$. First, he computes another public key $I = x_s H_p(P_s)$ called "key image", then applies the following transformations:

$$L_i \begin{cases} q_i G, & if \quad i = s \\ q_i G + w_i P_i, & if \quad i \neq s \end{cases}$$

$$R_i \begin{cases} q_i H_p(P_i), & if \ i = s \\ q_i H_p(P_i) + w_i I, & if \ i \neq s \end{cases}$$

$q_i$ and $w_i$ are random number selected from $[1,\ldots,l]$, then, compute the non-interactive challenge

$$c = H_s(m, L_1, \ldots, L_n, R_1, \ldots, R_n)$$

Finally, the signer compute

$$c_i \begin{cases} w_i, & if \ i \neq s \\ c - \sum_{i=0}^n c_i \ mod \ l, & if \ i = s \end{cases}$$

$$r_i \begin{cases} q_i, & if \ i \neq s \\ q_s - c_s x, & if \ i = s \end{cases}$$

The one-time ring signature is

$$\sigma = (I, c_1, \ldots, c_n, r_1, \ldots r_n)$$

- Verification

Any verifier can apply the transformation

$$\begin{cases} L_i^{'} = r_i G + c_i P_i \\ R_i^{'} = r_i H_p(P_i) + c_i I \end{cases}$$

then checks if the equation

$$\sum_{i=0}^n c_i = H_s(m, L_0^{'}, \ldots, L_n^{'}, R_0^{'}, \ldots R_n^{'}) \ \ holds.$$

*B. stealth address*

With aforementioned one-time ring signature, voters can cast their ballot among particular set. But there is still one more thing, the information on the ballot cannot be public until tally phase. Thus, the message must be send after encryption.

We apply the unlinkable payments scheme proposed by Nicolas van Saberhagen [ring-sig2], which allows sender to generate different destination base on the same public key. In the other word, once we encode the candidate into public key, voters can send transaction to different address even though they cast to the same candidate.

It means that voters can obfuscate the observers who they actually vote for. Stealth address can be incorporated into our system as following:

- A candidate $C_i$ ( or so called receiver ) has two standard elliptic key pair: $(a_i, A_i)$ , $(b_i, B_i)$ , where $A_i = a_i G$, $B_i = b_i G$.
- For a voter ( or so called sender ) who want to generate a stealth address for selected candidate $C_j$:

He chooses a random number $r \in [1, l-1]$ and compute $R = rG$, the corresponding stealth address $SA = H_s(rA_j)G + B_j$, then a ballot for the candidate $C_j$ is $(SA, R)$.

- Once the verifier get the private keys: $(a_j, b_j)$ of $C_j$ and the ballot information $(SA, R)$, he can compute

$$x = H_s(a_j R) + b_j$$
$$xG = (H_s(a_j R) + b_j)G$$

If the ballot is directed to corresponding $C_j$:

$$xG = (H_s(a_j R)G + b_j)G = H_s(a_j rG)G + b_j G$$
$$= H_s(rA_j)G + B_j = SA$$

However, there may be too many redundancies computation when verifying ballots. We can further simplify the computation of tally by sharing all candidates' first private key pair $(a_i, A_i)$ as $(a, A)$. then the ballot $(SA, R)$ is for the Candidate $C_j$ if

$$xG = H_s(rA)G + B_j = SA$$

There are other elliptic curve cryptography encryptions like integrated encryption scheme [ECC-IES] may be used, but must concern that the message (selected candidate) for encryption are limited and public-known.

*C. Key management*

In a general election, information about election results should not be opened until tally phase. We need more than one people to share the first private key of candidate $a$. We take Diffie- Hellman key exchange to solve the problem as following:

(1) Alice and Bob pick a random number $r_a, r_b$ and announce $r_a G, r_b G$ separately.
(2) Both them can compute $r_a r_b G$ and make it public
(3) After voting phase, they submit $r_a, r_b$ separately.

$(r_a r_b, r_a r_b G)$ will be candidates' first shared key pair. The method can be easily extended to multiparty, and ensure even only one of them doesn't collude, secret won't be opened early.

We can further import deposit scheme to make participants follow the protocol. The method we chosen doesn't needs the secret be generated before sharing to avoid the involvement of centralized authority. Furthermore, we can ask key managers to sign a message then publish in first step, ensuring that they have the corresponding private key of public key.

## III. PROPOSED SYSTEM

To make whole process transparent and convenient, we deploy our system on Ethereum smart contract and save all necessary information on it.

The election can be split into three parts:

### A. Setup

- Voter list: A set of public key, each one is a point on elliptic point and corresponds to a valid voter. The public key of valid voter should be appended on the voter list after authentication. The authentication process can be either centralized or decentralized. The former one can be conducted by third party. The system assures Anonymity of voters even relationship between identity and public key are publicly linkable. The property enable participants to supervise authentication process by checking identity of voter on the list. The latter one is integrated with decentralized verification[dec-aut], allow only the voters with specific credential publish their public key to list.

- Candidate: For each candidate can be string or other data type. In order to generate public key for stealth address, voter can encode them into coordinate on elliptic curve by [h2p1, h2p2]. As long as voters have consensus on the deterministic hash function, ballots will be counted correctly at tally phase. Note that we hash the representation of each candidate $C_j$ to get aforementioned $B_j$, the corresponding $b_j$ is unnecessary in tally phase.

- Secret sharing: The chosen key managers build shared secret as <chapter Key management mentioned> on the Ethereum smart contract. To ensure that key managers will submit their individual secret, they may need to pay an appropriate amount of Ether as deposit.

### B. Voting

For each valid voter has one of private keys corresponding to public keys of voter list and knows public information including candidates' first public key A, candidate list from Ethereum smart contract.

First, compute a ballot $(SA, R)$ for selected candidate by $(A, B_i)$. $B_i$ is the hash of candidate $C_i$, the point on elliptic curve.

Then, the ballot be signed with owned private key and one-time ring signature to get signed ballot $(m, \sigma)$, $m = (SA, R)$

Third, send signed ballot on smart contract with different Ethereum account. Note that if ballot sent by address corresponding to original private key directly, anonymity will be destroyed immediately.

Complex operations cost much on smart contract, so we reduce the computation as much as possible without affecting the correctness of results. We separate the verification of ring signature from voting phase.

As a result, everyone can try to send invalid or re-voting ballots, but those wrong ballots can always be detected during the tally phase. A rational attacker will not cost gas to make a useless attack.

### C. Tally

Key managers should open their individual secret for retrieve their deposit and no one can vote after voting phase. Then, people who interested in election can access all information and tally the result by themselves. It means that they don't need to trust authority or someone, and the privacy of every participant is still protected by property of ring signature in tally phase.

For each valid ballot should signed with one-time ring signature. The signed ballot $(m, \sigma)$ will be counted to candidate $C_i$ if

1. key image $I$ never shown before
2. $\sum_{i=0}^{n} c_i = H_s(m, L_0', \ldots, L_n', R_0', \ldots R_n')$
3. $H_s(aR)G + B_i = SA$, $a$ is product of key managers' secret.

Restricted by the upper bound of gas per block on Ethereum, we cannot count the ballots on smart contract. However, there will still be the same election result if everyone has consensus on counting method, which can be defined earlier on blockchain.

## IV. EXPERIMENT AND DISCUSSION

### A. Time analysis

Our measurements are performed on a MacBook Pro running OS X 10.12.5 equipped with 2 core, 2.7 GHz Intel Core i5 and 8 GB DDR3 RAM. For communication with the blockchain conveniently, we written our code in Javascript. As figure 1. shows, the time spent by voter is mainly to sign the ballot with ring signature, which is linear to ring size. And the verification time is almost the same as the time of genration, since the bottleneck of both is computation of ring signature, but it is still acceptable for voters to keep their anonymity.

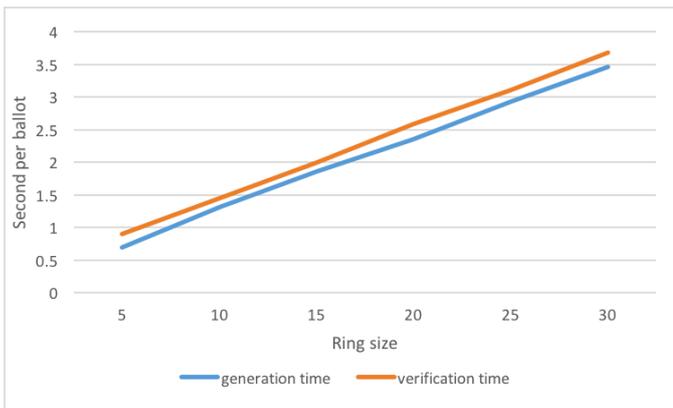

Fig. 1. The generation and verification time of a ballot.

*B. Gas cost*

It is obvious that data size of ballot is linear to the number of public key set, and the gas cost per voter is used to record information on the blockchain. Keep data on smart contract costs much more than keep on blockchain. In order to reduce voters' expenses as much as possible, we don't record the ballot information on Ethereum smart contract directly, but save it in normal transaction. The ballot information on smart contract is only transaction id of normal transaction, which allows us to reduce cost per voter significantly. We can further reduce that by save data on IPFS, which is a peer-to-peer distributed file system, then the ballots we save on the smart contract are pointer to particular file on IPFS. That is, the cost of each voter will be the lowest and constant, rather than linear to size of public key set. Namely, voters can choose the anonymity parameter they want without considering gas cost.

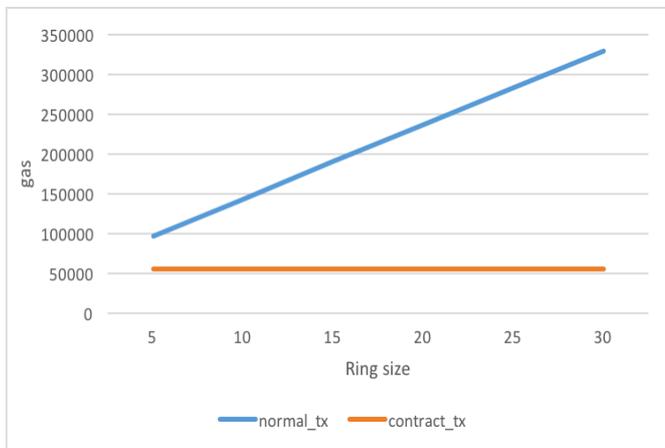

Fig2. Gas cost per ballot on Ethereum

*C. Security*

The security of our proposed system can be split into two parts: anonymity of voters and encryption of selected candidate. The former is protected by ring signature proposed by CryptoNote [ring-sig 2], which has been adopted by some cryptocurrency like Monero[XMR]. There are some studies analysis the traceability on Monero[traceablity1,2].

The temporal analysis doesn't work on our system because the transaction output is not related to input, and zero mix-ins problem can be eliminated by only tallying the ballot with more than specific number of ring size.

The encryption scheme we adopt performs a Diffie-Hellman exchange, which build a shared secret from selected random number and public key of recipient on elliptic curve(ECDH). Another risk is colluding of key managers, although we can increase difficulty by increasing number of key managers, it is still possible to be attacked. In the worst case, all of them collude to retrieve shared secret key, can only make them tally the result earlier. Neither they can destroy anonymity of voters nor change the result.

V. CONCLUSION

We proposed a decentralized anonymous voting system which only requires minimal trust in others and gas cost per voter. By putting all the information on Ethereum network, we make the whole election transparent and all participants have identical information. Since the tally phase could speed up via parallel computation, the system is also suitable for large-scale voting.

ACKNOWLEDGMENT *(Heading 5)*

The preferred spelling of the word "acknowledgment" in America is without an "e" after the "g." Avoid the stilted expression "one of us (R. B. G.) thanks ...". Instead, try "R. B. G. thanks...". Put sponsor acknowledgments in the unnumbered footnote on the first page.